# Klein-Gordon and square-root operator equations for two-spinors and scalars: perturbation calculations for hydrogen-like systems

Tobias Gleim

There exists a Klein-Gordon-like equation for a spin-1/2 particle in an electromagnetic field with 2-spinors as wave functions that is a direct consequence of the corresponding Dirac equation. Thus, it reproduces the same binding energies for an electron in a hydrogen atom as the Dirac equation. There is also a square-root equation for 2-spinors which can give the same binding energies up to the forth order of the fine structure constant, inclusively, what will be shown by means of a comparison of the non-relativistic limit plus a first relativistic correction of both equations for 2-spinors within the framework of a perturbation analysis. A parallel will also be drawn to the spin-0 case.

Square-root operator equations for spin-0 particles are already well known as a mean to circumvent the disadvantages of the Klein-Gordon equation (see e.g. [3]), that are e.g. a non-positive definite probability density and the simultaneous description of particles and anti-particles: both let appear rather doubtful a single-particle probability interpretation of the solutions of the Klein-Gordon equation. For spin-1/2 particles, the former drawback can be overcome by means of the Dirac equation, but the latter one only if one introduces an additional postulate: the so called "Dirac sea" (see e.g. [1,2,7]). But this cannot be applied to spin-0 particles. The same square-root operator equations may also emerge in the context of approximations of the Bethe-Salpeter equation (see e.g. [4]) when describing bound systems of fermion/ anti-fermion pairs. So there is enough motivation to be engaged in square-root operator equations for the spin-0 case.

But there also exists a pair of equations for spin-1/2 particles that consists of a Klein-Gordon-like equation – being a direct consequence of the Dirac equation – and a square-root equation, which both use 2-spinors as wave functions.

In the subsequent sections, these pairs of equations will be coupled to an electromagnetic field, but with the restriction to a Coulomb potential for a hydrogen or a hydrogen-like atomic system, respectively. For both pairs, the binding energies of the particle in that potential are known for the Klein-Gordon or Klein-Gordon-like equations, respectively (in case of spin-0 particles see e.g. [7] and with respect to the spin-1/2 particles see e.g. [6]). In the spin-1/2 case, this is a consequence of the fact that the Klein-Gordon-like equation follows directly from the corresponding Dirac equation.

We are able to compare these two equations with their square-root operator counterparts by means of a perturbation analysis for energy eigenvalues based on the expansion of their Hamiltonians in a small fine structure constant (emerging in the Coulomb potential of the hydrogen atom). The results of this comparison will be two pairs of Hamiltonians that can be split up into an unperturbed Schrödinger energy operator and a perturbation in the form of relativistic corrections to the latter non-relativistic Hamiltonian. The two Hamiltonians in each pair will be shown to coincide at least up to the forth order of the fine structure constant, inclusively. But this would mean that their energy eigenvalues for the bound states of the system under consideration would coincide up to the same order in this approximation, too: in this way, one would also obtain the approximated binding energies for the square-root operator equations of the hydrogen(-like) system.

Since the spin-0 case is slightly easier than the spin-1/2 case, we are going to start our investigations for the former particles and thus will be able to transfer many of the results to the spin-1/2 case in the section there after.

### *Square-root operator equation for spin-0 particles*

An eigenvalue equation based on the square-root operator equation for spin-0 particles of mass $m$ possesses the subsequent form (where we are going to use natural units, i.e. $\hbar$ and $c$ are set to one):

$$E\psi(\vec{x}) = \pm\left(\sqrt{m^2 + \hat{\vec{p}}^2}\psi\right)(\vec{x}) \tag{1}$$

with the momentum operator in position space representation, $\hat{\vec{p}} = -i\vec{\nabla}$.

Here one can see that there is a separate equation for particles (with positive energies) and antiparticles (with negative energies or vice versa). In that what follows, we restrict ourselves to the equation for particles. By means of a minimal coupling to an electromagnetic field with the potential energy $V$ and the vector potential $\vec{A}$, (1) turns into:

$$E\psi(\vec{x}) = \left(\sqrt{m^2 + (\hat{\vec{p}} - e\vec{A})^2}\psi\right)(\vec{x},t) + V\psi(\vec{x}) \tag{2}$$

Since it is an equation for spin-0 particles, the wave function $\psi$ must be a scalar.
If the spin-0 particle stays within the Coulomb potential of a hydrogen-atom-like system,

$$V(r) = -\frac{\alpha}{r} \tag{3}$$

(with the fine structure constant $\alpha = e^2 \approx 1/137$ and the elementary charge $e$), (2) simplifies to

$$E\psi(\vec{x}) = \left(\sqrt{m^2 + \hat{\vec{p}}^2}\psi\right)(\vec{x}) + V(r)\psi(\vec{x}). \tag{4}$$

A formal expansion of the square-root term in a power series up to the second order of the squared momentum operator according to the formula

$$\sqrt{1+x} = 1 + \frac{1}{2}x - \frac{1}{8}x^2 + \ldots, \tag{5 a}$$

with $x = \dfrac{\hat{\vec{p}}^2}{m}$ yields:

$$\sqrt{m^2 + \hat{\vec{p}}^2} \approx m + \frac{\hat{\vec{p}}^2}{2m} - \frac{1}{2m}\left(\frac{\hat{\vec{p}}^2}{2m}\right)^2 + \ldots \tag{5 b}$$

Thus, (4) becomes

$$E'\psi(\vec{x}) \approx \left[\frac{\hat{\vec{p}}^2}{2m} - \frac{1}{2m}\left(\frac{\hat{\vec{p}}^2}{2m}\right)^2\right]\psi(\vec{x}) + V(r)\psi(\vec{x}) \tag{6}$$

with $E' = E - m$. But this is the Schrödinger equation with Coulomb potential, wherein the non-relativistic operator of the kinetic energy has been supplemented by a first relativistic correction.
This result can be contrasted with a spin-0 particle in a Coulomb potential being described with the aid of an eigenvalue equation based on the Klein-Gordon equation:

$$(E - V(r))^2 \psi(\vec{x}) = (\hat{\vec{p}}^2 + m^2)\psi(\vec{x}). \tag{7}$$

This eigenvalue problem can be solved analytically in an exact way. It gives the subsequent binding energies (see e.g. [7, 8]):

$$E_{nl} = \frac{m}{\sqrt{1 + \dfrac{\alpha^2}{(n - \varepsilon_l)^2}}} \tag{8}$$

with

$$\varepsilon_l = l + \frac{1}{2} - \sqrt{\left(l + \frac{1}{2}\right)^2 - \alpha^2} \ . \tag{9}$$

Here, the energy $E_{nl}$ depends on the main quantum number $n = 1, 2, 3,...$ and the orbital quantum number $l = 0, 1, 2,..., n-1$

But equation (7) can be investigated in the non-relativistic limit with a first relativistic correction, too. To this end, we cast (7) into the following form:

$$E'\left(1 + \frac{E'}{2m}\right)\psi = \left[\frac{\hat{\vec{p}}^2}{2m} + \frac{V}{m}E' + V - \frac{V^2}{2m}\right]\psi \ . \tag{10}$$

It will be our aim to perform a perturbation calculation for eigenvalues with (6) and (10), respectively, in order to obtain the binding energies for the Coulomb potential problem. Therefore, we are going to start from the well-known non-relativstic wave functions $\psi_{nlm}$ (with the magnetic quantum number $m = -l, -l+1,..., l-1, l$) that are eigenfuctions of the Schrödinger energy operator

$$\hat{H}_0 = \frac{\hat{\vec{p}}^2}{2m} + V(r), \tag{11 a}$$

i.e.

$$\hat{H}_0 \psi_{nlm} = E_n \psi_{nlm}, \tag{11 b}$$

and that are describing the bound states, because we might expect that the relativistic eigenvalue problem can be approximated very well by the non-relativistic one (for the Klein-Gordon equation, this is known to be the case). The corresponding energy eigenvalues are the non-relativistic binding energies:

$$E_n = -\frac{m\alpha^2}{2n^2} \ . \tag{12}$$

But in (6), $\hat{H}_0$ is accompanied by a perturbation operator

$$\hat{H}_1 = -\frac{1}{2m}\left(\frac{\hat{\vec{p}}^2}{2m}\right)^2 \tag{13}$$

In order to take into account the influence of this perturbation operator on the binding energies of the Coulomb problem, one applies first order perturbation theory:

$$E_{nl} \approx \left\langle \psi_{nlm} \left| \hat{H}_0 + \hat{H}_1 \right| \psi_{nlm} \right\rangle = E_n + \Delta E_1, \tag{14}$$

with

$$\Delta E_1 = \left\langle \psi_{nlm} \left| \hat{H}_1 \right| \psi_{nlm} \right\rangle = \left\langle \hat{H}_1 \right\rangle. \tag{15}$$

The term $\Delta E_1$, being a correction to the non-relativistic energy $E_n$, can be determined in this approximation as an expectation value of $\hat{H}_1$ within the wave functions $\psi_{nlm}$. Therefore, it is useful to realise that expectation values of powers of the radius $r$ within the non-relativistic energy eigenfunctions $\psi_{nlm}$ of the hydrogen atom lead to the following proportionality relation (which is valid for positive as well as negative exponents $k$, see e.g. [6, 9]):

$$\left\langle r^k \right\rangle \propto \alpha^{-k}, \tag{16}$$

wherein $\alpha$ denotes the fine structure constant again. With (16), it is possible to determine the order in $\alpha$ of each term in (14) and (15), respectively. The Schrödinger energy $E_n$ (12), e.g., is of order $\alpha^2$, because the Coulomb potential as well as the operator $\hat{H}_0$ (11 a) are exactly of that order. The latter can be explained by the fact that the momentum operator $\hat{\vec{p}}$ possesses the dimension of an inverse length $1/r$ and therefore is of order $\alpha$. The perturbation operator $\hat{H}_1$ (13) in (6) is already of order $\alpha^4$. Up to this order in $\alpha$, i.e. $\alpha^4$, inclusively, we are going to approximate the reformulated Klein-Gordon equation (10). To this end, we are going to apply an iteration procedure: we replace $E'\psi$ in (10) by the right hand side of

$$E'_k \psi = \hat{H}(\alpha^k) \psi, \tag{17}$$

where $\hat{H}(\alpha^k)$ signifies a Hamiltonian only containing terms up to the order $\alpha^k$, inclusively. Multiplication of (10) from the left hand side by $\left(1 + \dfrac{\hat{H}(\alpha^k)}{2m}\right)^{-1}$ yields:

$$E'\psi = \left(1 + \frac{\hat{H}(\alpha^k)}{2m}\right)^{-1} \left[\frac{\hat{\vec{p}}^2}{2m} + \frac{V}{m}\hat{H}(\alpha^k) + V - \frac{V^2}{2m}\right]\psi. \tag{18}$$

That multiplication term can be expressed formally in a power series expansion:

$$(1+x)^{-1} \approx 1 - x + x^2 - \ldots \tag{19}$$

On the right hand side of (18), only terms up to the order of $\alpha^{k+1}$, inclusively, will be considered, i.e. all terms which according to (16), will come to be of that order or lower within the expectation value (14). On the other hand, all the terms of higher order will be neglected. With $E' = E'_{k+1}$, (18) gives the equation

$$E'_{k+1}\psi = \hat{H}(\alpha^{k+1})\psi, \tag{20}$$

which again can become the starting-point of the next iteration step.
The iteration may begin in order $\alpha^0$ with

$$\hat{H}(\alpha^0) = m \tag{21}$$

(here we restrict ourselves to particles again: due to this, we use $+m$ instead of $-m$ in (21)). In the next iteration step, i.e. up to the order $\alpha^2$, as expected, the Hamiltonian of the non-relativistic Schrödinger equation with Coulomb potential is reproduced:

$$\hat{H}(\alpha^2) = \hat{H}_0 = \frac{\hat{\vec{p}}^2}{2m} + V(r). \tag{22}$$

By means of (11) and (13), we get a Hamiltonian being valid up to the order of $\alpha^4$, inclusively,

$$\hat{H}(\alpha^4) = \hat{H}_0 + \hat{H}_1 + V\hat{T} - \hat{T}V, \tag{23}$$

wherein we have denoted the kinetic energy by an operator $\hat{T}$:

$$\hat{T} = \frac{\hat{\vec{p}}^2}{2m} = \hat{H}_0 - V. \tag{24}$$

Besides the energy operator $\hat{H}_0 + \hat{H}_1$ of (6), the latter being the non-relativistic limit of the square-root operator equation (4) plus a first relativistic correction to it, Hamiltonian (23) contains two extra terms that are – as $\hat{H}_1$ – of order $\alpha^4$. Calculating – in analogy to (14) – the expectation value of (23) within the states $|\psi_{nlm}\rangle$ of the non-relativistic Coulomb problem, by means of (24) and due to the Hermiticity of $\hat{H}_0$ and (11 b), we obtain for those two extra terms:

$$\langle \psi_{nlm}|V\hat{T} - \hat{T}V|\psi_{nlm}\rangle = \langle \psi_{nlm}|V\hat{H}_0 - \hat{H}_0 V|\psi_{nlm}\rangle = E_n \langle \psi_{nlm}|V - V|\psi_{nlm}\rangle = 0. \tag{25}$$

This means that one can calculate the binding energies with the help of (14) up to the forth order in $\alpha$, inclusively, within the framework of a perturbation analysis for both equations, the square-root equation (4) and the Klein-Gordon equation (7), in the non-relativistic limit plus a first relativistic correction. Surprisingly, these energies are identical for both equations up to that order of $\alpha$, and are derived in the described way in many books about quantum mechanics and atomic physics by means of

$$\left\langle \frac{1}{r} \right\rangle_{nlm} = \frac{1}{a_0 n^2} \tag{26}$$

and

$$\left\langle \frac{1}{r^2} \right\rangle_{nlm} = \frac{1}{a_0^2 n^3 (l+1/2)}. \tag{27}$$

They yield (see e.g. [9, 10]):

$$E_{nl} \approx E_n - E_n \frac{\alpha^2}{n^2} \left( \frac{3}{4} - \frac{n}{l+1/2} \right). \tag{28}$$

(28) can also be reproduced by a power series expansion of the exact energies $E_{nl}$ (8) of the Klein-Gordon equation up to the forth order of $\alpha$, inclusively.
But for even higher than the forth order in $\alpha$, especially for the square-root operator equation, one has to apply a different kind of perturbation theory than the one used here (see [5]).
In the subsequent section, we will try to find a square-root equation for spin-1/2 particles, too, in order to couple it to an electromagnetic field. Within the non-relativistic limit, the Pauli equation with the correct Landé factor 2 for electrons should result. In close analogy to the spin-0 case, we are going to calculate the binding energies for a spin-1/2 particle in a Coulomb potential up to the order $\alpha^4$, inclusively.

*Square-root equation for spin-1/2 particles*
For spin-1/2 particles, one could just repeat equation (1), but this time, the wave function is not a scalar but a 2-spinor. This means, for a Lorentz transformation $\Lambda$, one would get the transformed wave function:

$$U(\Lambda)\psi(x) = D^{-1}(\Lambda)\psi(\Lambda x) \tag{29}$$

with a 4-vector $x = (t, \vec{x})$ and the representation $D(\Lambda)$ (that is a $2\times 2$–matrix) of the Lorentz transformation.

By minimal coupling, one would yield (2) again, if the spin-1/2 particle interacted with an electromagnetic field. But a formal power series expansion of the square-root in (2) shows, that this would not reproduce the well-known Pauli equation (see e.g. [9]),

$$E'\psi(\vec{x},t) = \left( \frac{(\hat{\vec{p}} - e\vec{A})^2}{2m} - \frac{e}{2m}\vec{\sigma}\cdot\vec{B} + V \right)\psi(\vec{x},t) \tag{30}$$

with the magnetic induction $\vec{B} = \vec{\nabla}\times\vec{A}$ and the Pauli matrices $\vec{\sigma}$. Instead, one has to assume that the Hamiltonian $\hat{H}$ in the spin-1/2 version of (2) must rather take on the form

$$\hat{H} = \sqrt{m^2 + (\vec{\sigma}\cdot(\hat{\vec{p}} - e\vec{A}(\vec{x})))^2} + V = \sqrt{m^2 + (\hat{\vec{p}} - e\vec{A})^2 - e\vec{\sigma}\cdot\vec{B}} + V, \tag{31}$$

what can be shown with the aid of a power series expansion of the square-root (see (5a)) by applying

$$\sigma_k \sigma_j = \delta_{kj} + i\sum_{l=1}^{3} \varepsilon_{kjl}\sigma_l. \tag{32}$$

With vanishing vector potential $\vec{A}$, Hamiltonian (31) coincides with the one in (4) again, which can be used to obtain the binding energies (28) being valid up to the order $\alpha^4$, inclusively. But of course, (28) is not valid for spin-1/2 particles, because it is dependent of the quantum number $l$ of the orbital momentum instead of $j$ for the total angular momentum (with $|l - 1/2| \le j \le l + 1/2$). In the case of the Dirac equation (see e.g. [1, 2, 7, 9]), one gets instead of (8) and (9)

$$E_{nj} = \frac{m}{\sqrt{1 + \frac{\alpha^2}{(n - \varepsilon_j)^2}}} \tag{33}$$

with

$$\varepsilon_j = j + \frac{1}{2} - \sqrt{\left(j + \frac{1}{2}\right)^2 - \alpha^2} \tag{34}$$

and from this by a power series expansion up to the order $\alpha^4$, inclusively, one obtains

$$E_{nj} \approx E_n \left[ 1 - \frac{\alpha^2}{n^2}\left( \frac{3}{4} - \frac{n}{j + 1/2} \right) \right]. \tag{35}$$

In analogy to (14), (35) can be derived by means of a perturbation calculation with the subsequent Hamiltonian:

$$\hat{H} = \frac{1}{2m}\hat{\vec{p}}^2 + V - \frac{1}{2m}\left(\frac{\hat{\vec{p}}^2}{2m}\right)^2 - \frac{1}{8m^2}\hat{\vec{p}}^2 V + \frac{1}{4m^2}\vec{\sigma}\cdot\vec{\nabla}V\times\hat{\vec{p}}. \tag{36}$$

The Dirac equation with minimal coupling to an electromagnetic field takes on the form

$$\hat{E}\phi = \left(\vec{\alpha}\cdot\left(\hat{\vec{p}} - e\vec{A}\right) + \beta m + \mathbf{1}V\right)\phi, \tag{37 a}$$

wherein $\hat{E} = i\partial_t$, $\vec{\alpha}$ and $\beta$ denote the Dirac matrices and $\phi$ is a 4-spinor. By a multiplication from the left hand side with $\beta$ and with the help of the gamma matrices $\gamma^0 = \beta$ and $\vec{\gamma} = \beta\vec{\alpha}$, which can be combined to a 4-vector $(\gamma^\mu) = (\gamma^0, \vec{\gamma})$ with respect to a diagonal 4-metric $(g_{\mu\nu})_{\mu,\nu=0,1,2,3} = (g^{\mu\nu})_{\mu,\nu=0,1,2,3} = diag(1,-1,-1,-1)$ and which fulfil the anti-commutation relation $\gamma^\mu\gamma^\nu + \gamma^\nu\gamma^\mu = 2g^{\mu\nu}$, (37 a) can be converted into

$$0 = \left[\gamma^0\left(\hat{E} - V\right) - \vec{\gamma}\cdot\left(\hat{\vec{p}} - e\vec{A}\right) - m\right]\phi. \tag{37 b}$$

If one introduces here the vector potential $(A^\mu) = (A^0, \vec{A})$ with $V = eA^0$ as well as the operators $(\hat{p}^\mu) = (\hat{p}^0, \hat{\vec{p}})$ with $\hat{p}^0 = \hat{E}$ and $\hat{\pi}^\mu = \hat{p}^\mu$, (37 b) can be written in the compact form

$$0 = \left(\gamma^\mu\hat{\pi}_\mu - m\right)\phi, \tag{37 c}$$

where, due to Einstein's summation convention, an index appearing once in an upper and once in a lower position is automatically summed over from 0 to 3 and the 4-metric can be used to transform an upper index into a lower index (and vice versa): e.g. $\hat{\pi}_\mu = g_{\mu\nu}\hat{\pi}^\nu$.

When using the Weyl representation, the Dirac equation (37) can be cast into the form of a Klein-Gordon equation (for this, see especially [6]) with the aid of the following projection operators

$$P_\pm = \frac{1}{2}\left(1 \pm \gamma^5\right) = (P_\pm)^2, \quad P_\pm P_\mp = 0, \quad P_+ + P_- = 1, \tag{38}$$

with $\gamma^5 = i\gamma^0\gamma^1\gamma^2\gamma^3$. Besides of Hermiticity, they possess the properties

$$P_\pm \gamma^\mu = \gamma^\mu P_\mp,$$

which, when multiplied from the left hand side with $P_\mp$, yields due to (37c)

$$m\phi = (P_\pm + P_\mp)m\phi = \left(\gamma^\mu\hat{\pi}_\mu + m\right)P_\pm\phi$$

If we multiply this again from the left hand side with the operator $\left(\gamma^\mu\hat{\pi}_\mu - m\right)$ from the Dirac equation (37 c), then after some straightforward calculations follows the equation

$$\left(\hat{\pi}^\mu\hat{\pi}_\mu - m^2 - \tfrac{1}{2}e\sigma^{\mu\nu}F_{\mu\nu}\right)P_\pm\phi = 0 \tag{39}$$

with $\gamma^\mu\gamma^\nu = g^{\mu\nu} - i\sigma^{\mu\nu}$ and $F_{\mu\nu} = \partial_\mu A_\nu - \partial_\nu A_\mu$, containing among other things the operator $\hat{\pi}^\mu\hat{\pi}_\mu - m^2$ of the Klein-Gordon equation.

In the Weyl representation (symbolised by $(\bullet)_W$), the gamma matrices take on the form

$$\gamma^0 = \begin{pmatrix} 0 & 1_{2\times 2} \\ 1_{2\times 2} & 0 \end{pmatrix}, \quad \vec{\gamma} = \begin{pmatrix} 0 & -\vec{\sigma} \\ \vec{\sigma} & 0 \end{pmatrix}$$

with the 2×2 unity matrix $1_{2\times 2}$ and the Pauli matrices $\vec{\sigma}$. Within that representation, the projection operators $P_\pm$ with

$$(P_+)_W = \begin{pmatrix} 1_{2\times 2} & 0 \\ 0 & 0 \end{pmatrix}, \quad (P_-)_W = \begin{pmatrix} 0 & 0 \\ 0 & 1_{2\times 2} \end{pmatrix}$$

as well as the term $\sigma^{\mu\nu} F_{\mu\nu}$ take on a particularly simple form and (39) gives the subsequent two equations

$$(\hat{E} - V)^2 \psi_\pm = \left[ m^2 + (\hat{\vec{p}} - e\vec{A}(\vec{x}))^2 - e\vec{\sigma} \cdot (\vec{B} \mp i\vec{E}) \right] \psi_\pm, \tag{40}$$

wherein we have used the Lorentz gauge $\partial_\mu A^\mu = 0$ and $\vec{B}$ denotes the magnetic induction as well as $\vec{E}$ the electric field. In this equation, the wave function $\psi_\pm$ is a 2-spinor and can be derived from the 4-spinor $\phi$ by means of the projection operators in the Weyl representation:

$$\psi_\pm = (P_\pm \phi)_W. \tag{41}$$

Moreover, with the help of the charge conjugation operator can be shown that the wave functions $\psi_+$ and $\psi_-$ belong to particle and anti-particle solutions (or vice versa), respectively. In the following, we will restrict ourselves again to particle solutions and will abbreviate $\psi_+$ by $\psi$. Since (40) is an exact consequence of the Dirac equation (37), it does not surprise that (40) leads to the same binding energies (33) for the hydrogen problem (see [6]).

For a particle in a pure Coulomb potential $V(r)$ (i.e. with $\vec{A} \equiv 0$), the non-relativistic limit of (40) plus a first relativistic correction can be found by means of an iteration procedure in close analogy to (18) for the Klein-Gordon equation. The Hamiltonian $\hat{H}(\alpha^4)$ containing all terms up to the order of $\alpha^4$, can be represented in the same way as the one in (23). But this time, the perturbation operator is

$$\hat{H}_1 = -\frac{1}{2m} \left( \frac{\hat{\vec{p}}^2}{2m} \right)^2 \pm \frac{e}{2m} i\vec{\sigma} \cdot \vec{E} \tag{42}$$

with $e\vec{E} = -\vec{\nabla} V$. According to the preceding reflections, it must be able to reproduce the approximated binding energies $E_{nj}$ (35) up to the order of $\alpha^4$. This is astonishing, because (42) seems to be quite different from the perturbation operator that would arise from (36) (and that would act on a 2-spinor as well): the latter Hamiltonian comprises not only the relativistic correction terms of the kinetic energy, but also the so called spin-orbit coupling and the Darwin term (the last and last but one term in (36), respectively; see also [1, 2, 7, 9]).

Thus, it seems to be as if the following square-root operator equation for spin-1/2 particles with 2-spinors can reproduce the Pauli equation (30) as well as the binding energies $E_{nj}$ (35) of the hydrogen atom up to the order $\alpha^4$, inclusively, here presented as an eigenvalue equation:

$$(E - V)\psi = \sqrt{m^2 + (\hat{\vec{p}} - e\vec{A}(\vec{x}))^2 - e\vec{\sigma} \cdot (\vec{B} \mp i\vec{E})} \, \psi. \tag{43}$$

Therefore, with respect to the hydrogen atom and within the perturbation analysis we performed, there seem to be no experimental reason to reject (43). But there may be theoretical ones, e.g. the not evident relativistic invariance of (43). .